\documentclass[pra,twocolumn,floatfix,showpacs,superscriptaddress]{revtex4}
\usepackage{amssymb}
\usepackage{amsmath}
\usepackage{graphicx}
\usepackage{epsfig}
\usepackage{amsfonts}

\begin{document}

\title{Time-dependent Fr\"ohlich transformation approach for\\
two-atom entanglement generated by successive passage through a cavity}
\author{Yong Li}
\affiliation{Department of Physics and Astronomy, University of Basel, Klingelbergstrasse
82, 4056 Basel, Switzerland}
\author{C. Bruder}
\affiliation{Department of Physics and Astronomy, University of Basel, Klingelbergstrasse
82, 4056 Basel, Switzerland}
\author{C. P. Sun}
\affiliation{Institute of Theoretical Physics, Chinese Academy of Sciences, Beijing,
100080, China}
\date{\today }

\begin{abstract}
Time-dependent Fr\"ohlich transformations can be used to derive an effective
Hamiltonian for a class of quantum systems with time-dependent
perturbations. We use such a transformation for a system with time-dependent
atom-photon coupling induced by the classical motion of two atoms in an
inhomogeneous electromagnetic field. We calculate the entanglement between
the two atoms resulting from their motion through a cavity as a function of
their initial position difference and velocity.%
\end{abstract}

\maketitle


\section{Introduction}

Canonical transformations have been widely used in condensed matter physics
\cite{mah,wagner} to derive effective Hamiltonians by eliminating degrees of
freedom with low-energy excitations \cite{zh}. One of the most well-known
applications was Fr\"ohlich's derivation \cite{fro} of an attractive
electron-electron interaction from the original electron-phonon interaction.

Fr\"{o}hlich's approach (which was also studied by Nakajima \cite{Naka}) was
to apply a unitary transformation $H^{\prime }=\exp (-gS)H\exp (gS)$,
defined by an anti-Hermitian operator $S$, to the Hamiltonian $%
H=H^{0}+gH^{1} $, where $H^{0}$ describes noninteracting electrons and
phonons. His goal was to treat the electron-phonon interaction $gH^{1}$. The
small parameter $g $ is introduced to stress that the interaction
Hamiltonian is perturbative compared with the free Hamiltonian and can be
set to $1$ during/after the calculation. The transformation can be evaluated
order by order,%
\begin{align}
H^{\prime }& =H^{0}+gH^{1}+g[H^{0},S]+g^{2}[H^{1},S]  \notag \\
& +{\frac{g^{2}}{2}}[[H^{0},S],S]+O(g^{3}).
\end{align}
Fr\"{o}hlich eliminated the term linear in $g$ by requiring
\begin{equation}
H^{1}+[H^{0},S]=0  \label{elim}
\end{equation}%
to get the generator $S$. In the following, we will use the term \textit{Fr%
\"{o}hlich transformation} for a canonical transformation that fulfills Eq.~(%
\ref{elim}). The transformation leads to a phonon-induced interaction among
electrons with the potential $V=g^{2}\langle \lbrack H^{1},S]/2\rangle $ by
averaging over the low-energy phonon states. This effective Hamiltonian can
be attractive or repulsive with a singularity at the energy shell.

Since the interaction between atoms and electromagnetic field is similar to
the electron-phonon coupling, it is natural to try to use Fr\"{o}hlich
transformations in quantum optics \cite{sun}. In the large-detuning limit,
i.e., if the difference between the atomic level spacing and the frequency
of the light field is much larger than the coupling strength, the
singularity at the energy shell is avoided, and the Fr\"{o}hlich
transformation is expected to work well. It will result in an effective
Hamiltonian which can also be obtained from the adiabatic elimination method
\cite{eberly90} and is equivalent to generic second-order perturbation
theory.

However, in quantum optics and atomic physics, most realistic systems
involve time-dependent classical fields. Obviously, a time-independent Fr%
\"{o}hlich transformation cannot work for these cases. In this paper we use
an effective Hamiltonian approach to eliminate certain intermediate degrees
of freedom (e.g., the photon degree of freedom, or the atomic operators
concerning a certain atomic level) corresponding to the time-dependent terms
for the general case. A time-dependent transformation is used to make the
first-order terms zero and to keep the second-order terms. We will call this
method \emph{time-dependent Fr\"{o}hlich transformation (TDFT)}.

The paper is organized as follows: in Section \ref{sec:froehlich} we will
briefly introduce the TDFT to fix the notation. In Section \ref{sec:atoms}
and the following sections, we will use the TDFT approach to study the
creation of two-atom entanglement after two atoms successively pass a cavity
with a single mode field. As a result, we find that the entanglement depends
on the atomic velocity (i.e., the transit time) and the initial distance
between the two atoms. We determine the parameter regions for which the two
atoms are maximally entangled, and discuss the limits of applicability of
our method.

\section{Time-dependent canonical transformation}

\label{sec:froehlich} We consider a general time-dependent quantum system
with a Hamiltonian
\begin{equation}
H=H^{0}+H^{1}(t),  \label{hamil-gen}
\end{equation}
where the unperturbed part $H^{0}$ is time-independent, and $H^{1}(t)$ is a
time-dependent perturbation ($|H^{1}(t)|\ll|H^{0}|$). We perform the
following time-dependent transformation:
\begin{equation}
\left\vert \psi^{\prime}(t)\right\rangle =e^{-S(t)}\left\vert \psi
(t)\right\rangle ,
\end{equation}
where $|\psi(t)\rangle$ is a quantum state whose time dependence is governed
by the Hamiltonian $H$. The transformed state $|\psi^{\prime}(t)\rangle$
evolves according to the unitarily transformed Hamiltonian%
\begin{align}
H^{\prime} & =e^{-S}He^{S}+i\left( \partial_{t}e^{-S}\right) \cdot e^{S}
\notag \\
& =H+\sum_{n=1}^{\infty}\frac{1}{n!}[...[H,S],...,S]  \notag \\
& +i\partial_{t}\left( \sum_{n=0}^{\infty}\frac{(-S)^{n}}{n!}\right)
\cdot\left( \sum_{n=0}^{\infty}\frac{S^{n}}{n!}\right) .
\end{align}
Keeping the terms up to second order, we obtain the following effective
Hamiltonian:
\begin{align}
H_{\mathrm{eff}}^{\prime} & = H^{0}+(H^{1}+[H^{0},S]-i\partial _{t}S)  \notag
\\
& +\frac{1}{2}[(H^{1}+[H^{0},S]-i\partial_{t}S),S]+\frac{1}{2}[H^{1},S].
\end{align}
If the operator $S(t)$ in the time-dependent transformation is chosen such
as to make the first-order term of the effective Hamiltonian zero, that is%
\begin{equation}
H^{1}+[H^{0},S]-i\partial_{t}S=0,  \label{1 order}
\end{equation}
the effective Hamiltonian takes the simple form
\begin{equation}
H_{\mathrm{eff}}^{\prime}=H^{0}+\frac{1}{2}[H^{1},S].  \label{ef}
\end{equation}
The canonical transformation described by Eqs.~(\ref{1 order}) and (\ref{ef}%
) is the so-called \emph{time-dependent Fr\"{o}hlich transformation (TDFT)}.

The form of the effective Hamiltonian after TDFT is similar to that in the
time-independent case. Moreover, one also needs to have the formal solution
for Eq.~(\ref{1 order}) to give an explicit expression for the effective
Hamiltonian (\ref{ef}). We assume $\{|m\rangle $ $|m=0,1,2,...\}$ to be a
set of eigenstates for the time-independent zeroth-order Hamiltonian $H^{0}$%
, and $E_{m}$ to be the eigenvalue for $\left\vert m\right\rangle $. The
matrix elements of Eq.~(\ref{1 order}) in this basis lead to
\begin{equation}
H_{mn}^{1}+(E_{m}-E_{n})S_{mn}-i\partial _{t}S_{mn}=0,
\end{equation}%
where we have used the notation $(\partial _{t}S)_{mn}=\partial _{t}S_{mn}$
since $\left\vert m\right\rangle $ is time-independent. Thus, the solution
for the transformation matrix elements $S_{mn}$ is given by \cite{note1}
\begin{equation}
S(t)=-i\sum_{m,n}\int_{0}^{t}e^{-iE_{mn}(t-t^{\prime })}H_{mn}^{1}(t^{\prime
})\mathrm{d}t^{\prime }\left\vert m\right\rangle \left\langle n\right\vert
\label{s-form}
\end{equation}%
with $E_{mn}:=E_{m}-E_{n}$ for any $m$ and $n$.

We would like to remark that TDFT is equivalent to second-order
time-dependent perturbation theory (as proved in Appendix \ref{appendix}).
However, the effective Hamiltonian obtained after applying TDFT, Eq. (\ref%
{ef}), which contains only second-order interaction terms and no first-order
interaction terms, is usually more convenient to be evaluated than the
original Hamiltonian. For the time-independent weak-coupling atom-photon
system (e.g., the systems given in Refs. \cite{sun,Zheng and Guo}, or the
electron-phonon systems in Refs. \cite{zh,Naka}), time-independent
perturbation theory will give a second-order perturbation solution involving
an infinite-dimensional Hilbert space. The equivalent conventional Fr\"{o}%
hlich approach will give a decoupled effective Hamiltonian which can involve
only the atomic part and can reduce to an analytically-solvable Hamiltonian
in a finite-dimensional atomic Hilbert space. Similarly, in the
time-dependent case, the interaction terms are time-dependent in the
original Hamiltonian (\ref{hamil-gen}), and TDFT will give an effective
(time-dependent) Hamiltonian, which may be decoupled from the photonic part
and can be evaluated in the atomic Hilbert space. Therefore, applying TDFT
can be much more efficient than using time-dependent perturbation theory.

In the following sections, we will give an example for the power of TDFT,
namely, a system of two atoms that cross a single-mode optical cavity. This
is by no means the only example. We expect that many atom-light or
electron-phonon models that have been studied by the conventional Fr\"{o}%
hlich approach (or other equivalent approaches) to eliminate certain degrees
of freedom, are suitable to be evaluated by TDFT if the interaction depends
on time. To conclude: TDFT is powerful for atom-light or electron-phonon
models whose time-dependent interaction is perturbative.

\section{Time-dependent model for two atoms passing a cavity successively}

\label{sec:atoms} 
\begin{figure}[tbp]
\centerline{\includegraphics[width=8cm,height=5cm]{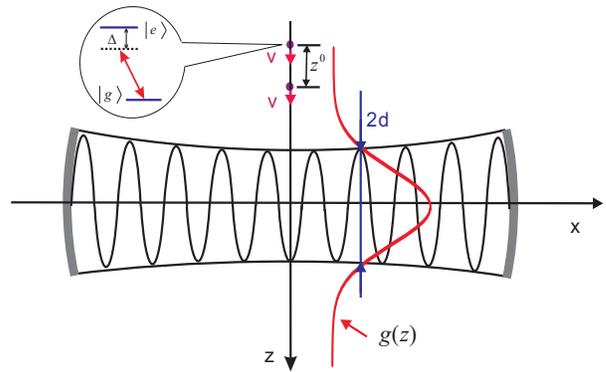}}
\caption{(Color online) Two atoms crossing the optical cavity. The width of
the cavity is $2d$ and determines the profile of the coupling constant $%
g(z)=g_{0}\exp (-z^{2}/d^{2})$.}
\label{a1}
\end{figure}

We now use the TDFT method developed above to study a
realistic physics problem in quantum optics. Entanglement is a defining
feature of quantum mechanics that has no classical counterpart. It is an
interesting issue to entangle two atoms separated by a large distance that
have no direct interaction. Numerous proposals have been made for entangling
atoms trapped in a cavity/cavities \cite%
{atom-cav01,atom-cav02,atom-cav03,Zheng and
Guo,Haroche01,HarocheRMP,Haroche02,atom-cav04,atom-cav05,Walther05,Walther06}%
. Here, we will propose a scheme to entangle two identical two-level atoms,
which are not trapped in a cavity but pass the single-mode optical cavity
sequentially in transverse direction, see Fig.~\ref{a1}. Our goal is to
calculate the degree of entanglement between these two atoms after this
process. The coupling of the atoms to the cavity field is
position-dependent, and their motion therefore causes a time-dependent
coupling. If the coupling energy is assumed to be much less than the
detuning between the atomic transition frequency and the optical frequency,
that is, the large-detuning condition is satisfied, we can use the TDFT to
get an effective time-dependent Hamiltonian, which only involves atom-atom
interaction terms and eliminates the optical field.

A similar idea to entangle two atoms crossing a far-off-resonant single-mode
cavity has been studied \cite{Zheng and Guo}. However, there are significant
differences between Ref. \cite{Zheng and Guo} and our model: (i) In the
proposal \cite{Zheng and Guo}, both two-level atoms enter (or leave) the
cavity with the same velocity at the same time. In our model, the two atoms
have different initial positions, i.e., they enter the cavity with the same
velocity at different times. In fact, we will study the degree of
entanglement between the atoms as a function of the difference in initial
position. (ii) In Ref. \cite{Zheng and Guo}, the coupling between the atoms
and the cavity is assumed to be constant. Therefore, it is possible to
obtain an effective Hamiltonian with a reduced atom-atom interaction
(assuming large detuning) by eliminating the photons (i.e., by means of the
conventional Fr\"{o}hlich transformation or other equivalent approaches). In
reality, the coupling depends on the position the atom and has a Gaussian
shape. Thus, the coupling is time-dependent when the atoms cross the cavity,
and this is our motivation to use TDFT to study the present model. There are
many other works on the generation of entanglement between two atoms
crossing a cavity. Reference \cite{HarocheRMP} describes experiments that
study entanglement between two atoms after crossing a resonant-coupling
cavity one by one. A two-qubit Grover quantum search algorithm is studied in
Ref. \cite{Haroche02} by looking at two atoms crossing a large-detuning
cavity (this model is similar to that in Ref. \cite{Zheng and Guo}). The
generation and purification of maximally entangled states of two $\Lambda $%
-type-atoms inside a large-detuning cavity have also been investigated \cite%
{Walther05}. The important difference between all of the above models and
our model is that the atom-photon coupling in the cavity is assumed to be
constant in these models but is time-dependent in our model.

As can be seen from Fig.~\ref{a1}, the atoms are assumed to have the same
constant velocity $v$ along the $z$-direction with both the initial
positions $z_{1}^{0}$ and $z_{2}^{0}$ far away from the cavity. The
Hamiltonian reads ($\hbar =1$)%
\begin{align}
H_{ori}& =\omega _{0}(\sigma _{1}^{z}+\sigma _{2}^{z})+\omega a^{\dag }a
\notag \\
& +g_{1}(t)\sigma _{1}^{+}a+g_{2}(t)\sigma _{2}^{+}a+h.c.,  \label{two-atom}
\end{align}%
where $\sigma _{j}^{+}=\left\vert e\left\rangle _{jj}\right\langle
g\right\vert $ and $\sigma _{j}^{z}=\left\vert e\left\rangle
_{jj}\right\langle e\right\vert $ with $\left\vert e\right\rangle $ and $%
\left\vert g\right\rangle $ the atomic excited and ground states; $%
g_{j}(t)=g(z_{j}^{0}+vt)\sin (kx)$ are the coupling constants which are
assumed to be of Gaussian form, $g(z)=g_{0}\exp (-z^{2}/d^{2})$, where $2d$
is the width of the cavity and $g_{0}$ the vacuum Rabi frequency. Here we
set $\sin (kx)=1$ since the atoms are assumed to pass the cavity in
transverse direction at a maximum of the standing light wave. It is also
assumed that the atomic velocity along the $x$- and $y$-direction is zero.
We ignore the back-action of the cavity to the momentum of the atoms since
we assume the velocity to be large, (e.g., $v\sim 10$ m/s), such that the
atomic decay time is long compared with the atomic cavity transit time. We
assume the condition of large detuning is also fulfilled, $|\Delta |\gg
|g_{0}|\gtrsim |g_{1,2}(t)|$ with the detuning $\Delta =\omega _{0}-\omega $%
, where $\omega _{0}$ ($\omega $) is the atomic transition frequency
(optical frequency).

In the interaction picture with respect to
\begin{equation}
H_{ori}^{0}=\omega (\sigma _{1}^{z}+\sigma _{2}^{z})+\omega a^{\dag }a,
\end{equation}%
the Hamiltonian (\ref{two-atom}) reads $H=H_{0}+H_{1}(t)$ where
\begin{align}
H_{0}& =\Delta (\sigma _{1}^{z}+\sigma _{2}^{z}),  \notag \\
H_{1}& =g_{1}(t)\sigma _{1}^{+}a+g_{2}(t)\sigma _{2}^{+}a+h.c..
\end{align}%
For large detuning, we can use a time-dependent Fr\"{o}hlich transformation
defined by
\begin{equation}
S(t)=x_{1}(t)\sigma _{1}^{+}a+x_{2}(t)\sigma _{2}^{+}a-h.c.
\end{equation}%
to eliminate the photon operators. The coefficients $x_{1,2}(t)$ satisfy the
following equations:
\begin{equation}
g_{j}(t)+\Delta x_{j}(t)-i\dot{x}_{j}(t)=0,\ \ \ (j=1,2)  \label{15}
\end{equation}%
according to Eq.~(\ref{1 order}). The explicit solution for the coefficients
$x_{1,2}(t)$ is given by \cite{note2}
\begin{equation}
x_{j}(t)=-i\int_{0}^{t}g_{j}(t^{\prime })e^{i\Delta (t^{\prime }-t)}\mathrm{d%
}t^{\prime },\ \ \ (j=1,2).  \label{coeffs}
\end{equation}

After the Fr\"{o}hlich transformation, the effective Hamiltonian reads (up
to a constant term)
\begin{align}
H_{\mathrm{eff}}^{\prime }(t)& =H_{0}+\frac{1}{2}[H_{1}(t),S(t)]  \notag \\
& =\Delta _{1}\sigma _{1}^{z}+\Delta _{2}\sigma _{2}^{z}+f\sigma
_{1}^{-}\sigma _{2}^{+}+f^{\ast }\sigma _{1}^{+}\sigma _{2}^{-},  \label{30}
\end{align}%
where
\begin{equation}
\Delta _{j}=\Delta -(g_{j}^{\ast }x_{j}+g_{j}x_{j}^{\ast })(1+2n_{p})
\end{equation}%
for $j=1,2$, $n_{p}=\langle a^{\dag }a\rangle $ is the mean photon number in
the cavity, and%
\begin{equation}
f(t)=-\frac{1}{2}(g_{1}^{\ast }x_{2}+g_{2}x_{1}^{\ast })
\end{equation}%
is the effective coupling between the two atoms. The (first-order) weak
interaction terms between the photons and atoms have been eliminated, and
only the induced (second-order) atom-atom interaction terms remain. The
Hamiltonian (\ref{30}), which contains no photon operators (except the
time-independent expectation value of the photon number operator), is much
simpler than the original one (\ref{two-atom}). It can be diagonalized in
the atomic basis
\begin{equation}
\{|g_{1}g_{2}\rangle ,|g_{1}e_{2}\rangle ,|e_{1}g_{2}\rangle
,|e_{1}e_{2}\rangle \}\equiv \{|gg\rangle ,|ge\rangle ,|eg\rangle
,|ee\rangle \}.
\end{equation}

So far, we do not need any condition except the large detuning to get the
simple effective Hamiltonian (\ref{30}). In the following section, we will
further simplify the effective Hamiltonian (\ref{30}) using typical
parameters for the cavity-QED system. As we will see below, the adiabatic
condition is satisfied since the time-dependent couplings $g_{1,2}(t)$\ are
slowly varying (that is, $|\dot{g}_{j}(t)/g_{j}(t)|$\ $\ll $\ $|\Delta |$
for $j=1,2$). Of course, if the adiabatic condition is satisfied, the result
obtained by TDFT in the next section can also be obtained by using adiabatic
elimination. However, the elimination approach cannot be expected to
describe our model if the adiabatic condition is not fulfilled.

\section{Dynamical generation of two-atom entanglement}

In the following, we will simplify the Hamiltonian (\ref{30}) further by
taking into account typical parameters for the cavity-QED system.

First, we will rewrite the coefficients (\ref{coeffs}) as follows:
\begin{align}
x_{j}(t)& \simeq -ie^{-i\Delta t}\int_{-\infty }^{t}g_{j}(t^{\prime
})e^{i\Delta t^{\prime }}\mathrm{d}t^{\prime }  \notag \\
& \simeq \frac{-e^{-i\Delta t}}{\Delta }\left\{ g_{j}(t)e^{i\Delta t}+\frac{%
2v}{d}\int_{-\infty }^{t}g_{j}(t^{\prime })e^{i\Delta t^{\prime }}\mathrm{d}%
t^{\prime }\right\}  \notag \\
& =-\frac{g_{j}(t)}{\Delta }-\frac{2iv}{\Delta d}x_{j}(t).  \label{xt}
\end{align}%
Here, we have used the explicit Gaussian form of the coupling constants. We
have also used
\begin{equation}
\int_{-\infty }^{t}(z_{j}^{0}+vt)g_{j}(t^{\prime })e^{i\Delta t^{\prime }}%
\mathrm{d}t^{\prime }\rightarrow d\int_{-\infty }^{t}g_{j}(t^{\prime
})e^{i\Delta t^{\prime }}\mathrm{d}t^{\prime }
\end{equation}%
since $z_{j}^{0}+vt\sim d$ in the effective integration range and the change
of $g_{j}(t)$ (also change of $z_{j}^{0}+vt$) is much slower than that of $%
e^{i\Delta t}$:
\begin{equation}
F(g_{j}(t))\sim F(z_{j}^{0}+vt)\sim \frac{v}{d},\text{ }F(e^{-i\Delta
t})\sim \Delta ,  \label{sysparameters}
\end{equation}%
where $F(x):=\partial _{t}x/x\ $denotes the change of $x$. For typical
system parameters \cite{kimble,blais,you} (wavelength of the atomic
transition $\lambda =1000$ nm, detuning $\Delta =10^{4}$ MHz, $g_{0}=100$
MHz, cavity width (in the $z$-direction) $d=30\lambda $, and atomic velocity
of the order of $v=10$ m/s), $(v/d)$ $\sim $ $10^{-1}$ MHz $\ll |\Delta
|=10^{4}$ MHz, that is, the atomic cavity transit time $t_{transit}=2d/v$ is
much larger than the time $1/\Delta $ according to the detuning. Actually,
the fact that $v/d\ll |\Delta |$\ (that is, $|\dot{g}_{j}(t)/g_{j}(t)|$\ $%
\ll $\ $|\Delta |$ for $j=1,2$) means the adiabatic condition is satisfied.

Using these parameter values, $x_{j}(t)$ in Eq.~(\ref{xt}) can be further
simplified as
\begin{equation*}
x_{j}(t)\simeq-\frac{g_{j}(t)}{\Delta}-10^{-5}ix_{j}(t)\simeq-\frac{g_{j}(t)%
}{\Delta}=x_{j}^{\ast}(t).
\end{equation*}

The coefficients in Eq.~(\ref{30}) thus take the following form:
\begin{equation}
\Delta _{j}=\Delta +\frac{2(1+2n_{p})g_{j}^{2}(t)}{\Delta },
\label{sim-delta}
\end{equation}%
and the effective coupling is
\begin{equation}
f(t)=\frac{g_{1}(t)g_{2}(t)}{\Delta }.  \label{sim-f}
\end{equation}

We will now further assume that the shift terms in $\Delta_{j}$ can be
ignored since $2g_{j}^{2}(t)/|\Delta|\ll|\Delta|$, i.e., we will replace the
coefficients $\Delta_{j}$ by $\Delta$. Hence,
\begin{equation}
H_{\mathrm{eff}}^{\prime}(t)=\Delta\sigma_{1}^{z}+\Delta\sigma_{2}^{z}+f%
\sigma_{1}^{-}\sigma_{2}^{+}+f\sigma_{1}^{+}\sigma_{2}^{-}.  \label{eff}
\end{equation}
We now apply the transformation $U(t)=\exp\{(\Delta\sigma_{1}^{z}+\Delta
\sigma_{2}^{z})t\}$ to the Hamiltonian $H_{\mathrm{eff}}^{\prime}(t)$
defined in Eq. (\ref{eff}) and obtain
\begin{equation}
H_{\mathrm{eff}}^{\prime\prime}(t)=f\sigma_{1}^{-}\sigma_{2}^{+}+f\sigma
_{1}^{+}\sigma_{2}^{-}.  \label{hamil-eff}
\end{equation}
The time evolution of a general state governed by $H_{\mathrm{eff}%
}^{\prime\prime}(t)$ is given as%
\begin{equation}
\left\vert \psi(t)\right\rangle =C_{gg}(t)\left\vert gg\right\rangle
+C_{ge}(t)\left\vert ge\right\rangle +C_{eg}(t)\left\vert eg\right\rangle
+C_{ee}(t)\left\vert ee\right\rangle ,
\end{equation}
where $C_{gg}(t)\equiv C_{gg}(0)$, $C_{ee}(t)\equiv C_{ee}(0)$, and
\begin{align}
C_{ge}(t) & =C_{ge}(0)\cos\theta(t)-iC_{eg}(0)\sin\theta(t), \\
C_{eg}(t) & =-iC_{ge}(0)\sin\theta(t)+C_{eg}(0)\cos\theta(t),
\end{align}
with $\theta(t)=\int_{0}^{t}f(t^{\prime})\mathrm{d}t^{\prime}\simeq
\int_{-\infty}^{t}f(t^{\prime})\mathrm{d}t^{\prime}$. Hence, the system
described by Eq.~(\ref{hamil-eff}) is characterized by a closed subspace $%
\{|ge\rangle,|eg\rangle\}$.

In what follows, we will investigate what degree of entanglement can be
obtained after both atoms pass the cavity in transverse direction. We assume
that the two atoms are prepared in the initial state $|\psi (0)\rangle
=|ge\rangle $. The time evolution generated by Eq.~(\ref{hamil-eff}) leads
to
\begin{equation}
|\psi (t)\rangle =\cos \theta (t)|ge\rangle -i\sin \theta (t)|eg\rangle .
\label{ft}
\end{equation}%
After both atoms have passed through the cavity and are far outside, that is
$t\rightarrow +\infty $, one has
\begin{equation}
\theta (+\infty )=\sqrt{\frac{\pi }{2}}\frac{g_{0}^{2}d}{v\Delta }\mathrm{%
\exp }\left[ -\frac{(z^{0})^{2}}{2d^{2}}\right] ,
\end{equation}%
where $z^{0}=z_{1}^{0}-z_{2}^{0}$ denotes the difference between the initial
atomic positions.

In general, the state shown in Eq. (\ref{ft}) corresponds to an entangled
state of the two atoms. In the following, we will study the degree of
entanglement as a function of $z^{0}$ and the atomic velocity $v$ using the
entanglement entropy that is defined as
\begin{equation}
E(\left\vert \psi(t)\right\rangle )=-\mathrm{Tr}(\rho_{1}\log_{2}%
\rho_{1}(t)).
\end{equation}
Here, $|\psi(t)\rangle$ is a pure state, and $\rho_{1}(t)=\mathrm{Tr}%
_{2}(\left\vert \psi(t)\right\rangle \left\langle \psi(t)\right\vert )$ is
the reduced density matrix of the first atom. Evaluating this expression for
the state shown in Eq.~(\ref{ft}), we obtain
\begin{align}
E(\left\vert \psi(+\infty)\right\rangle ) & =-\cos^{2}\left[ \theta (+\infty)%
\right] \log_{2}\cos^{2}\left[ \theta(+\infty)\right]  \notag \\
& -\sin^{2}\left[ \theta(+\infty)\right] \log_{2}\sin^{2}\left[
\theta(+\infty)\right] .  \label{entanglement}
\end{align}

\begin{figure}[ptb]
\centerline{\includegraphics[width=7cm,height=5cm]{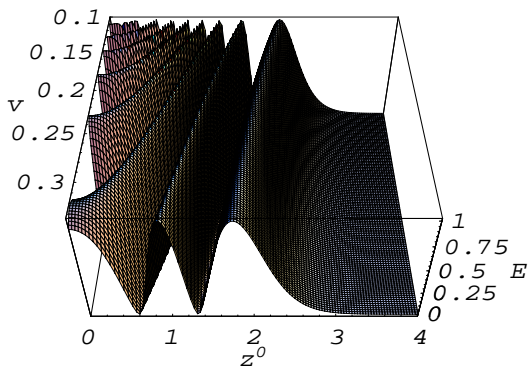}}
\caption{(Color online) Entanglement entropy $E$ as a function of atomic
velocity $v$ (in units of $g_{0}^{2}d/\Delta$) and initial atomic position
difference $z^{0}$ (in units of $d$).}
\label{a2}
\end{figure}

\begin{figure}[ptb]
\centerline{\includegraphics[width=7cm,height=5cm]{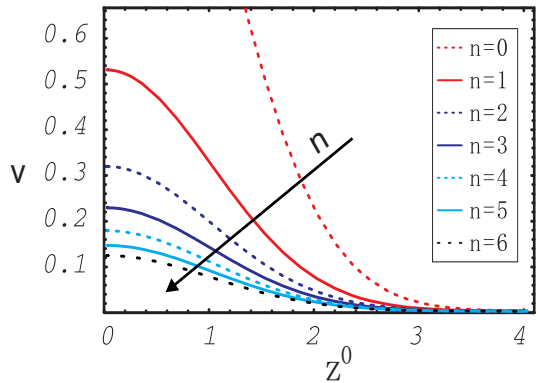}}
\caption{(Color online) Relation between atomic velocity $v$ (in units of $%
g_{0}^{2}d/\Delta$) and initial atomic position difference $z^{0}$ (in units
of $d$) along the lines of maximal entanglement $E(+\infty)=1$, or $\protect%
\theta (+\infty)=(2n+1)\protect\pi/4$, for $n=0,1,...,6$. }
\label{a3}
\end{figure}

A maximally entangled state for the atoms occurs for $\theta(+\infty
)=(2n+1)\pi/4$ for any integer $n$. Figure~\ref{a2} shows the entanglement
entropy $E(|\psi(+\infty)\rangle)$ as a function of initial atomic position
difference $z^{0}$ (in units of $d$) and atomic velocity $v$ (in units of $%
g_{0}^{2}d/\Delta$, which is 30 m/s for the system parameters discussed
after Eq.~(\ref{sysparameters})). The possible values for $z^{0}$ and $v$
which make the entanglement maximal (that is, $E(|\psi(+\infty)\rangle)=1$)
are shown in Fig.~\ref{a3}.

It is interesting that there may be maximally entangled states for the two
atoms even if the first atom has left the cavity before the second atom
begins to enter it. For example, when the velocity $v$ is smaller than about
$0.23g_{0}^{2}d/\Delta $ (corresponding to about $7$ m/s for the system
parameters discussed after Eq.~(\ref{sysparameters})), it is possible to
obtain a maximally entangled state although the initial atomic position
difference $z^{0}$ can be larger than $2d$ (i.e., the approximate transverse
width of cavity), see Fig.\ref{a3}.

In the above calculation, we have ignored the effect of (photonic and
atomic) decay and photonic back-action to the atoms for the following
reasons: (i) we assume that the velocity is of the order of $10$ m/s. Higher
velocities will lead to a reduced degree of entanglement; lower velocities
will lead to optical back-action to the atoms and photonic/atomic decay.
(ii) we have assumed large detuning to reduce the atomic decay. (iii) the
mean number of photons $n_{p}$ is assumed to be small (e.g. $10^{-4}$) in
order to reduce the effect due to photonic decay. For example, the atomic
cavity transit time (which is close to the interaction time) is less than
the effective photonic decoherence time: $2d/v\sim 10^{-5}$ s $%
<1/(n_{p}\kappa )\sim 10^{-4}$ s (where $\kappa \sim 10^{8}$ Hz is the
typical cavity one-photon damping rate \cite{blais}). Also, we have chosen $%
\Delta _{1,2}(t) $ $\rightarrow $ $\Delta $ which amounts to ignoring the
effect of the shift terms in Eq.~(\ref{sim-delta}). The explicit form of $%
\Delta _{1,2}(t)$ shows that if the initial atomic position difference $%
z^{0} $ is close to $0$, then $\Delta _{1}(t)\rightarrow \Delta _{2}(t)$,
and the effect of the shift terms can be neglected.

Finally, we would like to remark that the time-dependent Fr\"{o}hlich
transformation is valid under the condition of weak coupling (which is
equivalent to the condition of large detuning here): $H_{1}(t)\ll H_{0}$ ($%
g_{0}\ll \Delta $). The adiabatic condition that the atomic cavity transit
time $t_{transit}=2d/v$ is much larger than $1/\Delta $, is only used to
simplify the effective coupling in Eq.~(\ref{30}) for the present realistic
atoms-cavity system. Actually, this adiabatic condition is independent of
the above large detuning condition for the time-dependent Fr\"{o}hlich
transformation. The above argument proves that, the TDFT approach is valid
independent of whether the adiabatic condition is fulfilled or not.

\section{Conclusion}

Using the time-dependent Fr\"{o}hlich transformation, we have calculated the
degree of atomic entanglement between two atoms passing an optical cavity
sequentially. The Fr\"{o}hlich transformation eliminates the photonic
operators and induces an effective atom-atom interaction. We have determined
the velocities and initial atomic position differences for which the
entanglement is maximal, and we have shown that there may be maximally
entangled states for the two atoms even if the first atom has left the
cavity before the second atom begins to enter it.

A number of time-dependent canonical transformation methods have been
proposed \cite{wagner,wang} to discuss time-dependent problems. The
difference of our approach to a general time-dependent canonical
transformation is that our TDFT requires the generator $S(t)$ in the
canonical transformation to satisfy Eq.~(\ref{1 order}) in order to
eliminate some intermediate degree of freedom for a general system. As an
example, we have studied a system of two atoms that pass a cavity, which can
be described by an effective atom-atom interaction Hamiltonian by using the
TDFT to eliminate the photon degrees of freedom. In fact, the TDFT can be
used to treat certain time-dependent atom-light systems by eliminating some
of the atomic degree of freedom (e.g., one atomic level) like the Fr\"{o}%
hlich transformation for the time-independent cases. The TDFT presented here
works well not only for optical and atomic systems, but also in the field of
condensed matter physics (e.g., time-dependent electron-phonon interactions)
or other systems with weak (first-order) time-dependent interactions.

This work was supported by the European Union under contract IST-3-015708-IP
EuroSQIP, by the Swiss NSF, and the NCCR Nanoscience. It was also supported
by the NSFC with grant Nos. 90203018, 10574133, 10474104 and 60433050, and
NFRPC with No. 2005CB724508.

\appendix

\section{Equivalence to time-dependent second-order perturbation theory}

\label{appendix} To check the range of validity of the above effective
Hamiltonian obtained by TDFT, we compare with the results of standard
time-dependent perturbation theory up to of second order.

To apply time-dependent second-order perturbation theory for the Hamiltonian
$H=H^{0}+H^{1}(t)$ given in Eq. (\ref{hamil-gen}), we assume $\{|m\rangle$ $%
|m=0,1,2...\}$ to be a complete set of eigenstates for the time-independent
zeroth-order Hamiltonian $H^{0}$, with eigenvalues $E_{m}$. A state $%
|\psi(t)\rangle$ satisfies the Schr\"{o}dinger equation%
\begin{equation}
i\partial_{t}\left\vert \psi(t)\right\rangle =[H^{0}+H^{1}(t)]\left\vert
\psi(t)\right\rangle .  \label{schrodiner-eq}
\end{equation}

We now do perturbation theory in $H^{1}(t)$ by assuming
\begin{equation}
\left\vert \psi(t)\right\rangle \equiv\sum_{l=0}\left\vert
\psi^{(l)}(t)\right\rangle
=\sum_{n=0,l=0}C_{n}^{(l)}(t)e^{-iE_{n}t}\left\vert n\right\rangle .
\label{expand-order}
\end{equation}
Here, $|\psi^{(l)}(t)\rangle$ is the $l$-order contribution to the
perturbation expansion. Replacing $\left\vert \psi(t)\right\rangle $ in Eq. (%
\ref{schrodiner-eq}) by the perturbation expansion and comparing the
coefficients order by order, we obtain the following equation:%
\begin{equation}
i\dot{C}_{m}^{(l+1)}(t)=\sum_{n}C_{n}^{(l)}(t)e^{iE_{mn}t}H_{mn}^{1}(t).
\label{coefficent-equation}
\end{equation}

If the initial state is $\left\vert k\right\rangle $, the zeroth-order
solution for the coefficients $C_{n}(t)$ is
\begin{equation}
C_{m}^{(0)}(t)=\delta _{mk}.
\end{equation}%
According to Eq.~(\ref{coefficent-equation}) the first-order solution is
\begin{equation}
C_{m}^{(1)}(t)=-i\int_{0}^{t}e^{iE_{mk}t^{\prime }}H_{mk}^{1}(t^{\prime })%
\mathrm{d}t^{\prime }.  \label{1-order perturb}
\end{equation}%
The second-order solution is
\begin{equation}
C_{m}^{(2)}(t)=-\sum_{n}\int_{0}^{t}\mathrm{d}t^{\prime }e^{iE_{mn}t^{\prime
}}H_{mn}^{1}(t^{\prime })\int_{0}^{t^{\prime }}\mathrm{d}t^{\prime \prime
}e^{iE_{nk}t^{\prime \prime }}H_{nk}^{1}(t^{\prime \prime }).
\label{2-order perturb}
\end{equation}

We will now consider the time evolution of the state predicted by the TDFT
method. It follows from the Schr\"{o}dinger equation
\begin{equation}
i\partial_{t}\left\vert \psi^{\prime}(t)\right\rangle =(H^{0}+\frac{1}{2}%
[H^{1},S])\left\vert \psi^{\prime}(t)\right\rangle
\end{equation}
governed by the effective Hamiltonian $H_{\mathrm{eff}}^{\prime}$ in (\ref%
{ef}) after applying the TDFT method to the original Hamiltonian $%
H=H^{0}+H^{1}(t)$.

The connection to the original picture before the TDFT, i.e., the state $%
\left\vert \psi(t)\right\rangle $, whose time dependence is governed by $H$,
reads
\begin{equation}
\left\vert \psi(t)\right\rangle \equiv(1+S+\frac{1}{2}S^{2}+...)|\psi^{%
\prime }(t)\rangle.  \label{expand}
\end{equation}
Like in Eq.~(\ref{expand-order}), we can write down the perturbation
expansion for $\psi^{\prime}(t)$:
\begin{align}
\left\vert \psi^{\prime}(t)\right\rangle & \equiv\sum_{l=0}|\psi^{\prime
(l)}(t)\rangle=\sum_{n=0,l=0}C_{n}^{\prime(l)}(t)e^{-iE_{n}t}|n\rangle.
\end{align}

Starting with the initial state $\left\vert k\right\rangle $, and keeping
terms up to the zeroth-order in Eq.~(\ref{expand}), it is obvious that
\begin{equation}
|\psi^{(0)}(t)\rangle=|\psi^{\prime(0)}(t)\rangle,
\end{equation}
and
\begin{equation}
C_{m}^{(0)}(t)=C_{m}^{\prime(0)}(t)=\delta_{mk}.
\end{equation}
Since the perturbation term in $H_{\mathrm{eff}}^{\prime}$ is of second
order, the first-order correction for $\left\vert
\psi^{\prime}(t)\right\rangle $ vanishes: $\left\vert
\psi^{\prime(1)}(t)\right\rangle =0$. Then from Eq. (\ref{expand}), the
first-order correction for the state $|\psi(t)\rangle$ is
\begin{equation}
\left\vert \psi^{(1)}(t)\right\rangle =S\left\vert
\psi^{\prime(0)}(t)\right\rangle .
\end{equation}
Correspondingly, the first-order correction coefficients are%
\begin{equation}
C_{m}^{(1)}(t)=-i\int_{0}^{t}e^{iE_{mk}t^{\prime}}H_{mk}^{1}(t^{\prime })%
\mathrm{d}t^{\prime}.  \label{1-order Frohlich}
\end{equation}
The results of Eq.~(\ref{1-order Frohlich}) are equivalent to the standard
time-dependent perturbation theory as shown in Eq. (\ref{1-order perturb}).

The second-order correction for $\left\vert \psi (t)\right\rangle $ is%
\begin{equation}
\left\vert \psi ^{(2)}(t)\right\rangle =\left\vert \psi ^{\prime
(2)}(t)\right\rangle +\frac{1}{2}S^{2}|\psi ^{\prime (0)}(t)\rangle ,
\end{equation}%
where $|\psi ^{\prime (2)}(t)\rangle $ is the correction according to $%
[H^{1},S]/2$ in $H_{\mathrm{eff}}^{\prime }$ (\ref{ef}) with the exact form
being given as (according to Eq.~(\ref{1-order perturb}))
\begin{equation}
\left\vert \psi ^{\prime (2)}(t)\right\rangle =\sum_{m}C_{m}^{\prime
(2)}(t)e^{-iE_{m}t}\left\vert m\right\rangle
\end{equation}%
with%
\begin{equation*}
C_{m}^{\prime (2)}(t)=-\frac{i}{2}\int_{0}^{t}e^{iE_{mk}t^{\prime
}}[H^{1}(t^{\prime }),S(t^{\prime })]_{mk}\mathrm{d}t^{\prime }.
\end{equation*}%
Then the coefficients for the second-order corrections of the state $%
\left\vert \psi ^{(2)}(t)\right\rangle $ are given as%
\begin{equation*}
C_{m}^{(2)}(t)=C_{m}^{\prime (2)}(t)+\frac{1}{2}\sum_{n}S_{mn}S_{nk}.
\end{equation*}%
By replacing the operator $S(t)$ by that in Eq. (\ref{s-form}) in the above
equation, we obtain
\begin{eqnarray}
C_{m}^{(2)}(t) &=&-\sum_{n}\int_{0}^{t}\mathrm{d}t^{\prime
}e^{iE_{mn}t^{\prime }}H_{mn}^{1}(t^{\prime })  \notag \\
&&\times \int_{0}^{t^{\prime }}\mathrm{d}t^{\prime \prime
}e^{iE_{nk}t^{\prime \prime }}H_{nk}^{1}(t^{\prime \prime }),
\label{2-order Frohlich}
\end{eqnarray}%
which agrees with the expression given in Eq. (\ref{2-order perturb}) by
time-dependent second-order perturbation theory.

In this appendix, by reformulating the well-known perturbation solution to
the time-dependent Schr\"{o}dinger equation in our notations, we have shown
explicitly that the second-order time-dependent perturbation theory agrees
with TDFT.


\begin{thebibliography}{99}
\bibitem{mah} G.D. Mahan, \textit{Many-Particle Physics} (Plenum Press, New
York, 1990).

\bibitem{wagner} M. Wagner, \textit{Unitary Transformations in Solid-State
Physics} (North-Holland, Amsterdam, 1986).

\bibitem{zh} H. Zheng, M. Avignon, and K.H. Bennemann, Phys. Rev. B \textbf{%
49}, 9763 (1994); H. Zheng, \textit{ibid.} \textbf{50}, 6717 (1994); H.
Zheng and S.Y. Zhu, \textit{ibid.} \textbf{55}, 3803 (1997).

\bibitem{fro} H. Fr\"ohlich, Phys. Rev. \textbf{79}, 845 (1950); Proc. Roy.
Soc. \textbf{A215}, 291 (1952); Adv. Phys. \textbf{3}, 325 (1954).

\bibitem{Naka} S. Nakajima, Adv. Phys. \textbf{4}, 463 (1953).

\bibitem{sun} H.B. Zhu and C.P. Sun, Chinese Science (A) \textbf{30}, 928
(2000); Progress in Chinese Science \textbf{10}, 698 (2000).

\bibitem{eberly90} C.W. Gardiner, Phys. Rev. A \textbf{29}, 2814 (1984);
C.C. Gerry and J.H. Eberly, \textit{ibid}. \textbf{42}, 6805 (1990).

\bibitem{note1} Strictly speaking, there may be terms of the form $%
\sum_{m,n}u_{mn}e^{-iE_{mn}t}\left\vert m\right\rangle \left\langle
n\right\vert $ in the solution of $x_{j}(t)$ in Eq.~(\ref{s-form}) with $%
u_{mn}$ ($=-u_{nm}^{\ast}$) arbitrary constants.

\bibitem{atom-cav01} J.I. Cirac, P. Zoller, H.J. Kimble, H. Mabuchi, Phys.
Rev. Lett. \textbf{78}, 3221 (1997); S.J. vanEnk, J.I. Cirac, P. Zoller,
\textit{ibid}. \textbf{78}, 4293 (1997); S.J. van Enk, J.I. Cirac, P.
Zoller, \textit{ibid}. \textbf{79}, 5178 (1997).

\bibitem{atom-cav02} T. Pellizzari, Phys. Rev. Lett. \textbf{79}, 5242
(1997).

\bibitem{atom-cav03} S. Bose, P. L. Knight, M. B. Plenio, and V. Vedral,
Phys. Rev. Lett. \textbf{83}, 5158 (1999)%
; M. B. Plenio, S. F. Huelga, A. Beige, and P. L. Knight, Phys. Rev. A
\textbf{59}, 2468 (1999)
.

\bibitem{Zheng and Guo} S.-B. Zheng and G.-C. Guo, Phys. Rev. Lett. \textbf{%
85}, 2392 (2000).%

\bibitem{Haroche01} A. Rauschenbeutel, P. Bertet, S. Osnaghi, G. Nogues, M.
Brune, J.M. Raimond, and S. Haroche, Phys. Rev. A \textbf{64}, 050301(R)
(2001).%

\bibitem{HarocheRMP} J.M. Raimond, M. Brune, and S. Haroche, Rev. Mod. Phys.
\textbf{73}, 565 (2001).%

\bibitem{Haroche02} F. Yamaguchi, P. Milman, M. Brune, J.M. Raimond, and S.
Haroche, Phy. Rev. A \textbf{66}, 010302(R) (2002).%

\bibitem{atom-cav04} L.-M. Duan and H.J. Kimble, Phys. Rev. Lett. \textbf{90}%
, 253601 (2003).

\bibitem{atom-cav05} S. Mancini and S. Bose, Phys. Rev. A \textbf{70},
022307 (2004).

\bibitem{Walther05} P. Lougovski, E. Solano, and H. Walther, Phy. Rev. A
\textbf{71}, 013811 (2005).%

\bibitem{Walther06} S. Rinner, E. Werner, T. Becker and H. Walther, Phys.
Rev. A \textbf{74}, 041802(R) (2006).%

\bibitem{note2} Strictly speaking, there may be terms of the form $u_{j}%
\mathrm{e}^{-i\Delta t}$ in the solution of $x_{j}(t)$ in Eq.~(\ref{coeffs})
(for $j=1,2$) where $u_{j}$ are arbitrary constants.

\bibitem{kimble} C.J. Hood, T.W. Lynn, A.C. Doherty, A.S. Parkins, and H.J.
Kimble, Science \textbf{287}, 1447 (2000).

\bibitem{blais} A. Blais, R.-S. Huang, A. Wallraff, S.M. Girvin, and R.J.
Schoelkopf, Phys. Rev. A \textbf{69}, 062320 (2004).

\bibitem{you} P. Zhang, Y. Li, C.P. Sun, and L. You, Phys. Rev. A \textbf{70}%
, 063804 (2004).

\bibitem{wang} L.-X. Cen, X.Q. Li, Y.J. Yan, H.Z. Zheng, and S.J. Wang,
Phys. Rev. Lett. \textbf{90}, 147902 (2003); Li-Xiang Cen, Z.D. Wang, and
S.J. Wang, Phys. Rev. A \textbf{74}, 032321 (2006).
\end{thebibliography}
\end{document}